\documentclass[11pt]{article}
\usepackage{moriond,epsfig, xspace}

\pdfoutput=1
\def\NIMA{{\em Nucl.\,Instrum.\,Methods}\,A}

\def\PLB{{\em Phys.\,Lett.}\,B}
\def\PRL{{\em Phys.\,Rev.\,Lett.}}
\def\PRD{{\em Phys.\,Rev.}\,D}

\def\JINST{{\em JINST}}
\def\JPG{{\em J.\,Phys.}\,G}

\def\be{\begin{equation}}
\def\ee{\end{equation}}
\def\bea{\begin{eqnarray}}
\def\eea{\end{eqnarray}}

\newcommand{\CL}{C.L.\ }

\newcommand{\CLs}{\ensuremath{\textrm{CL}_{\textrm{s}}}\xspace}

\newcommand{\pp}{\ensuremath{pp}\xspace}

\newcommand{\Bs}{\ensuremath{B^0_s}\xspace}
\newcommand{\Bd}{\ensuremath{B^0_d}\xspace}

\newcommand{\Bsmumu}{\ensuremath{\Bs\to\mu^+\mu^-}\xspace}
\newcommand{\Bdmumu}{\ensuremath{\Bd\to\mu^+\mu^-}\xspace}
\newcommand{\Bmumu}{\ensuremath{B^0\to\mu^+\mu^-}\xspace}

\newcommand{\Bmm}{\ensuremath{B^0_{(s)}\to \mu^+\mu^-}\xspace}

\newcommand{\Jpsi}{\ensuremath{J\!/\!\psi}\xspace}

\newcommand{\BRof}[1]{\ensuremath{{\cal B}(#1)}\xspace}

\newcommand{\gev}{\ensuremath{\,{\rm GeV}}\xspace}
\newcommand{\tev}{\ensuremath{\,{\rm TeV}}\xspace}

\newcommand{\gevc}{\ensuremath{\,{\rm GeV}\!/\!c}\xspace}
\newcommand{\mevcc}{\ensuremath{\,{\rm MeV}\!/\!c^2}\xspace}

\newcommand{\invfb}{\ensuremath{\,{\rm fb}^{-1}}\xspace}
\newcommand{\invpb}{\ensuremath{\,{\rm pb}^{-1}}\xspace}

\newcommand{\Enow}{\ensuremath{\,{\sqrt{s}=7\tev}}\xspace}

\newcommand{\PT}{\ensuremath{P_{\rm T}}\xspace}
\newcommand{\ET}{\ensuremath{E_{\rm T}}\xspace}

\begin{document}
\vspace*{4cm}
\title{\boldmath Search for \Bsmumu and \Bdmumu at LHCb}

\author{M.-O. Bettler}

\address{Istituto Nazionale di Fisica Nucleare (INFN), Sezione Firenze, Italy\\ and\\ 
European Organization for Nuclear Research (CERN), Geneva, Switzerland}

\maketitle\abstracts{
A search for the decays \Bsmumu and \Bdmumu is performed with about 37\invpb of \pp collisions at \Enow collected by the LHCb experiment at the Large Hadron Collider at CERN.
The observed numbers of events are consistent with the background expectations. 
The resulting upper limits on the branching fractions are \BRof\Bsmumu $< 5.6 \times 10^{-8}$ and \BRof\Bmumu $<1.5 \times 10^{-8}$ at 95\% confidence level.}

%==========================
\section{Introduction}
\label{sec:introduction}
%==========================

Within the Standard Model (SM) the \Bdmumu and \Bsmumu flavour changing neutral current transitions are rare as they occur only via loop diagrams and are helicity suppressed~\footnote{In this proceedings the inclusion of charge-conjugate states is implicit.}. 
Their branching fractions (BF) predicted by the SM~\cite{Buras2010} are $\BRof\Bsmumu = (0.32 \pm 0.02)\times 10^{-8}$ and $\BRof\Bmumu  = (0.010 \pm 0.001) \times 10^{-8}$.

In New Physics models, their BF can be significantly enhanced, although in some models they are lowered.
For instance, within the Minimal Supersymmetric SM~\cite{MSSM}, the BF can get contributions $\propto \tan^6\beta/M_A^4$, where $M_A$ denotes the pseudoscalar Higgs mass and $\tan\beta$ the ratio of Higgs vacuum expectation values.  

The best published $95\%$ \CL limits come from the D0 collaboration~\cite{d0_PLB} ($6.1~\invfb$), \BRof\Bsmumu $<5.1 \times 10^{-8}$, and from the CDF collaboration~\cite{cdf_PRL} ($2~\invfb$), \BRof\Bsmumu $<5.8 \times 10^{-8}$ and \BRof\Bmumu $<1.8 \times 10^{-8}$. 
CDF has also presented preliminary results~\cite{cdf_preliminary} with $3.7~\invfb$, that lower the limits to \BRof\Bsmumu $<4.3 \times 10^{-8}$ and \BRof\Bmumu $<0.76 \times 10^{-8}$.

The measurements presented in these proceedings use about 37\invpb of integrated luminosity collected by LHCb between July and October 2010 at \Enow.

%==========================
\section{The LHCb detector}
\label{sec:detector}
%==========================

The LHCb detector~\cite{LHCbdetector} is a single-arm forward spectrometer with an angular coverage from about 10\,mrad to 300\,mrad in the bending plane.
It consists of a vertex locator, a warm dipole magnet, a tracking system, two RICH detectors, a calorimeter system and a muon system. 

Track momenta are measured to a precision of $\delta p / p = 0.35 \ (0.5)\,\%$ at 5 (100)\gevc. 
The RICH system provides charged hadron identification in a momentum range 2--100\gevc.
Typically, kaon identification efficiencies of over 90\% can be attained for a $\pi \to K$ fake rate below 10\%.
The calorimeter system identifies high transverse energy ($E_{\rm T}$) hadron, electron and photon candidates and provides information for the trigger.
The muon system provides information for the trigger and muon identification with an efficiency of $\sim 95$\,\% for a misidentification rate of about 1--2\,\% for momenta above 10\gevc.

LHCb has a two-level flexible and efficient trigger system exploiting the finite lifetime and large mass of heavy flavour hadrons to distinguish them from the dominant light quark processes.
The first trigger level is implemented in hardware while the second trigger level is software implemented. 
The forward geometry allows the LHCb first level trigger to collect events with one or two muons with \PT values as low as 1.4\gevc for single muon and $\PT(\mu_1)>0.48$\gevc and  $\PT(\mu_2) > 0.56$\gevc  for dimuon triggers. 
The $\ET$ threshold for the hadron trigger varied in the range 2.6 to 3.6\gev.

%The dimuon trigger line requires vertexed muon pairs of opposite charge with invariant mass $M_{\mu\mu} > 4.7$\gevcc.   
%A second trigger line, primarily to select \Jpsimumu events, requires $2.97 < M_{\mu\mu} < 3.21$\gevcc. 
%The remaining region of the dimuon invariant mass is also covered by trigger lines that in addition require the dimuon secondary vertex to be well separated from the primary vertex.
%Other trigger lines select generic displaced vertices, providing a high efficiency for purely hadronic decays.

\section{Analysis Strategy}

The search for \Bmm  at LHCb is described in detail in Ref.~\cite{paper}.
Most of the background is removed by selection cuts, keeping $\sim60$\,\% of the reconstructed signal decays.
Then each event is given a probability to be signal or background in a two-dimensional probability space defined by the dimuon invariant mass and a multivariate discriminant likelihood, the Geometrical Likelihood (GL)~\cite{Karlen,thesis2010_068}. 
The compatibility of the observed distribution of events with a given BF hypothesis is computed using the \CLs method~\cite{Read_02}. 
The number of expected signal events is evaluated by normalizing with channels of known BF to ensure that knowledge of the absolute luminosity and $b \bar{b}$ production cross-section is not required.

\subsection{Event selection}

The selection consists of loose requirements on track separation from the interaction point, decay vertex quality and compatibility of the reconstructed origin of the $B$ meson with the interaction point. 
The selection cuts were defined in simulation before starting data analysis.
Events passing the selection are considered  $B^0_{(s)} \to \mu^+ \mu^-$ candidates if their invariant mass lies  within 60\mevcc of the nominal $B^0_{(s)}$ mass. 
A similar selection is applied to the normalization channels, in order to minimize systematic errors in the ratio of efficiencies.
Assuming the BF predicted by the SM, 0.04 (0.3) \Bmm events are expected after all selection requirements. 
There are 343 (342) \Bmm candidates selected from data in the $B^0_s\ (B^0)$ mass window.

The dominant background after the \Bmm  selection is expected to be $b\bar{b}\to \mu \mu X$~\cite{roadmap}. 
This is confirmed by comparing the expected yield and the kinematic distributions of the sideband data with a $b\bar{b}\to \mu \mu X$ MC sample.
The muon misidentification probability as a function of momentum obtained from data using $K^0_S \to \pi^+ \pi^-$, $\Lambda \to p \pi^{-} $ and $\phi \to K^+K^-$ decays is in good agreement with MC expectations. 
It was found that the background from misidentified $B^0_{s,d}\to h^+h^{'-}$  is negligible for the amount of data used in this analysis.

\subsection{Signal and background likelihoods}
\label{sec:likelihoods}

The discrimination of the signal from the background is achieved through the combination of two independent variables: the GL and the invariant mass. 
The invariant mass in the search regions ($\pm 60\mevcc$ around the $B^0_{(s)}$  masses) is divided into six equal-width bins, and the GL into four equal-width bins distributed between zero and one. 

The GL combines information related with the topology and kinematics of the event as the $B^0_{(s)}$ lifetime, the minimum impact parameter of the two muons, the distance of closest approach of the two tracks, the $B^0_{(s)}$ impact parameter  and $p_T$ and the isolation of the muons with respect to the other tracks of the event.
These variables are combined using the method described in Ref.~\cite{Karlen,thesis2010_068}. 
The expected GL distribution for signal events is flat, while for background events it falls exponentially.

The GL variable is defined using MC simulations but calibrated with data using $B^0_{(s)}\to h^+h^{'-}$ selected as the signal events and triggered independently on the signal in order to avoid trigger bias.
The number of $B^0_{(s)}\to h^+h^{'-}$ events in each GL bin is obtained from a fit to the inclusive mass distribution.

The \Bmm  mass resolution is estimated from data via two methods.
The first uses an interpolation to $M_{\Bs}$ between the measured resolutions for $c\overline{c}$ resonances (\Jpsi, $\psi$(2S)) and $b\overline{b}$ resonances ($\Upsilon(1S)$, $\Upsilon(2S)$, $\Upsilon(3S)$) decaying into two muons. 
Both methods yield consistent results and their weighted average, $\sigma = 26.7 \pm 0.9\mevcc$, is taken as the invariant mass resolution for both $B^0$ and $B^0_{s}$ decays. 

The prediction of the number of background events in the signal regions is obtained by fitting an exponential function to the $\mu\mu$ mass sidebands independently in each GL bin.
The mass sidebands are defined in the range between $M_{B^0_{(s)}} \pm 600~(1200)$\mevcc for the lower (upper) two GL bins, excluding the two search regions ($M_{B^0_{(s)}} \pm 60$\mevcc).

\subsection{Normalization}
\label{sec:norm_factors}

The number of expected signal events is evaluated by normalizing with three channels of known BF, $B^+\to \Jpsi K^+$, $B^0_s\to \Jpsi \phi$ and $B^0 \to K^+\pi^-$.
The first two decays have similar trigger and muon identification efficiency to the signal but a different number of final-state particles, while the third channel has the same two-body topology but cannot be efficiently selected with the muon triggers.

For a given  normalization channel, \BRof\Bmm can be calculated as:
\begin{eqnarray}
\BRof\Bmm  &=& {\mathcal{B}_{\rm norm}}\times\frac{\rm
\epsilon_{norm}^{REC}
\epsilon_{norm}^{SEL|REC}
\epsilon_{norm}^{TRIG|SEL}
}{\rm
\epsilon_{sig}^{REC}
\epsilon_{sig}^{SEL|REC}
\epsilon_{sig}^{TRIG|SEL}
}\times\frac{f_{\rm norm}}{f_{B^0_{(s)}}}
\times\frac{N_{B^0_{(s)} \to \mu\mu}}{N_{\rm norm}} \nonumber\\
&=& \alpha_{B^0_{(s)} \to \mu\mu} \times N_{B^0_{(s)} \to \mu\mu}\,, \nonumber
\label{eq:normAlpha}
\end{eqnarray}
where $\alpha_{B^0_{(s)} \to \mu\mu}$ denotes the normalization factor,
$f_{B^0_{(s)}}$  the probability that a $b$-quark fragments 
into a $B^0_{(s)}$ and $f_{\rm norm}$ the probability that a $b$-quark fragments into the $b$-hadron relevant for the chosen normalization channel with BF $\mathcal{B}_{\rm norm}$. 
The reconstruction efficiency ($\epsilon^{\rm REC}$) includes the acceptance and particle identification,
$\epsilon^{\rm SEL|REC}$ denotes the selection efficiency on reconstructed events and  $\epsilon^{\rm TRIG|SEL}$ is the trigger efficiency on selected events.
The ratios of reconstruction and selection efficiencies are estimated from simulations and checked with data, while the ratios of trigger efficiencies on selected events are determined from data~\cite{TisTos}.

The normalization factors calculated using the three channels give compatible results; 
the final normalization factors are weighted averages and read $\alpha_{B^0_{(s)} \to \mu\mu}= (8.6 \pm 1.1) \times 10^{-9}$ and $\alpha_{B^0 \to \mu\mu}= (2.24 \pm 0.16) \times 10^{-9}$.

%==========================
\section{Results}
\label{sec:results}
%==========================

For each of the 24 bins, the expected number of background events is computed from the fits to the invariant mass sidebands.  
For a given BF hypothesis, the expected number of signal events is computed using the normalization factors and the signal likelihoods. 
For each BF hypothesis, the compatibility of the expected distributions with the observed distribution is evaluated using the \CLs method~\cite{Read_02}. 

\begin{figure}[t]
\centering
\begin{minipage}[b]{0.46\linewidth}
\includegraphics[width=\textwidth]{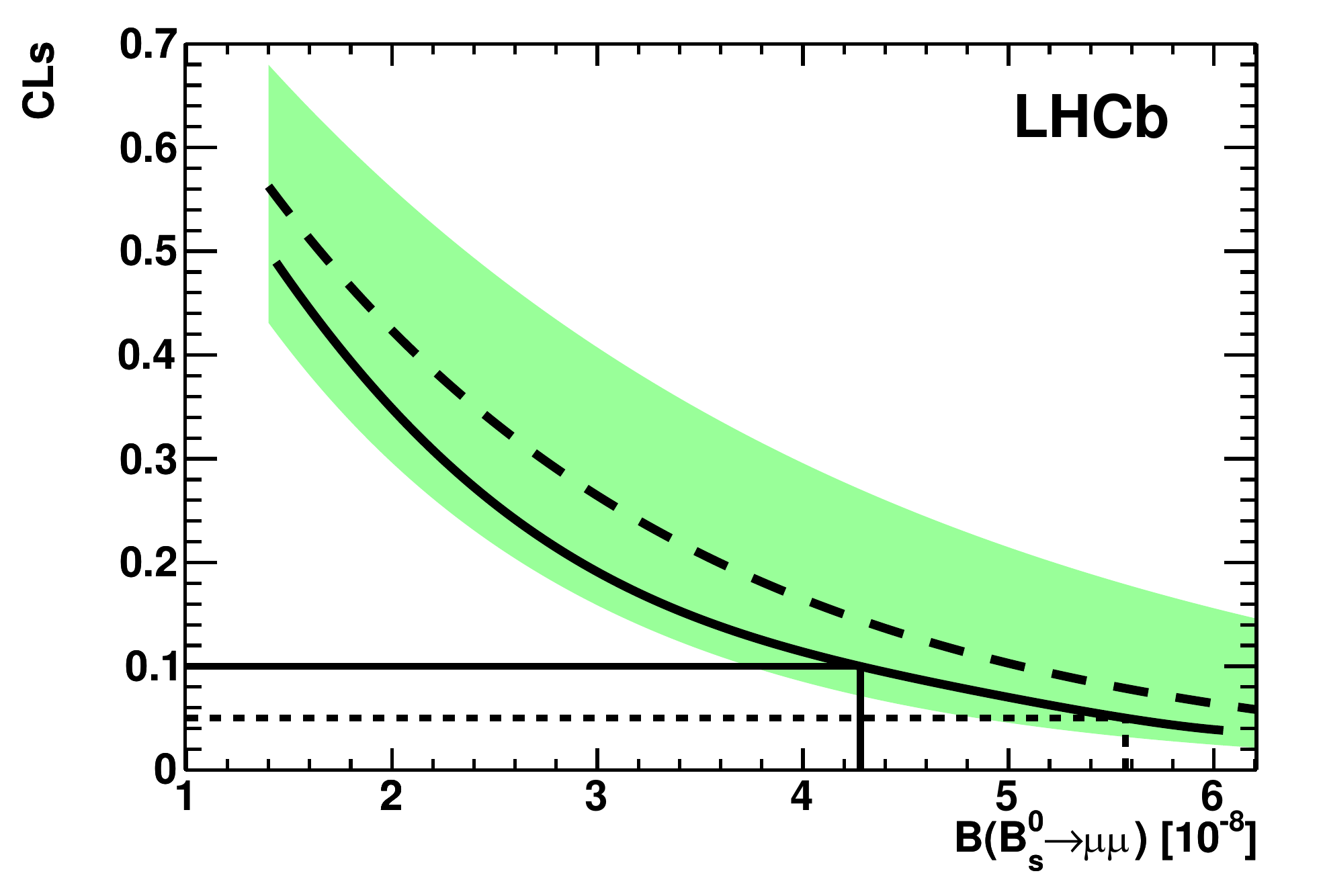}
\hspace{-77mm}\raisebox{2mm}{\textbf{(a)}}
\end{minipage}
\hfill
\begin{minipage}[b]{0.46\linewidth}
\includegraphics[width=\textwidth]{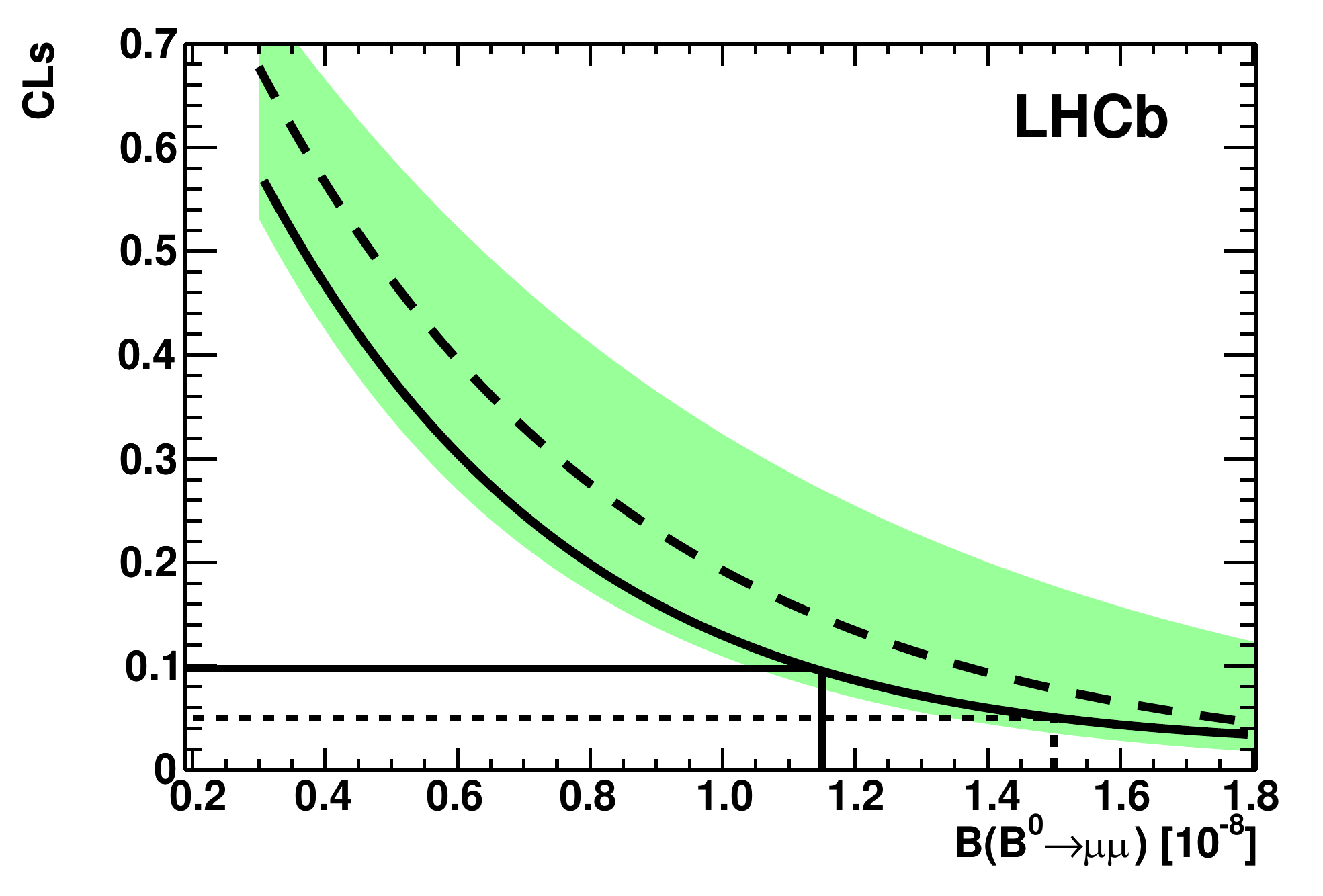}
\hspace{-77mm}\raisebox{2mm}{\textbf{(b)}}
\end{minipage}
\caption
{(a) Observed (solid curve) and expected (dashed curve) CL$_s$ values as a function of \BRof\Bsmumu.
The green shaded area contains the $\pm 1 \sigma$ interval of possible results compatible with the expected value when only background
is observed. The 90\,\% (95\,\%) \CL  observed value is identified by the solid (dashed) line. 
(b) the same for \BRof\Bmumu.}
\label{fig:CLsvsBR}
\end{figure}

The observed distribution of \CLs as a function of BF hypothesis can be seen in Fig.~\ref{fig:CLsvsBR}. 
The expected distributions of possible values of \CLs assuming the background-only hypothesis are also shown in the same figure as a green area covering the region of $\pm 1 \sigma$ of background-only compatible observations. 
The uncertainties in the signal and background likelihoods and 
normalization factors are used to compute the uncertainties in the background 
and signal predictions.
The upper limits read:
\begin{eqnarray}
\BRof\Bsmumu &<& 4.3~(5.6)\times10^{-8}~{\rm at}~90\,\% ~(95\,\%)~{\rm C.L.,}  \nonumber \\
\BRof\Bmumu  &<& 1.2~(1.5)\times10^{-8}~{\rm at}~90\,\% ~(95\,\%)~{\rm C.L.,} \nonumber
\end{eqnarray} 
while the expected values of the limits are $\BRof\Bsmumu < 5.1~(6.5)\times10^{-8}$  and $\BRof\Bmumu < 1.4~(1.8) \times 10^{-8}~{\rm at}~90\,\% ~(95\,\%)$ C.L.
The limits observed are similar to the best published limits for the \Bsmumu decay  and more restrictive for the \Bmumu  decay.

\section*{References}
\begin{thebibliography}{99}

\bibitem{Buras2010} 
A.J.~Buras,  %``Minimal flavour violation and beyond: Towards a flavour code for short distance dynamics'', 
arXiv:1012.1447; \\
E.~Gamiz {\it et al}, %``Neutral B meson mixing in unquenched lattice QCD'', 
\PRD \textbf{80} (2009) 014503; \\
A.J.\ Buras, %``Relations between $\Delta M_{s,d}$ and $B_{s,d} \to \mu^+ \mu^-$ in models with Minimal Flavour Violation'',
\PLB \textbf{566} (2003) 115.

\bibitem{MSSM} 
%L.~J.~Hall, R.~Rattazzi and U.~Sarid, %``The top quark mass in supersymmetric SO(10) unification'' 
%\PRD \textbf{50} (1994) 7048; \\ 
C. Hamzaoui, M. Pospelov and M. Toharia, %
``Higgs-mediated FCNC in supersymmetric models with large $\tan \beta$'',
\PRD \textbf{59} (1999) 095005;
S.Rai Choudhury and N. Gaur, \PLB \textbf{451} (1999) 86;
K.S. Babu and C.F. Kolda, %``Higgs-mediated $B_{s,d} \to \mu^+\mu^-$ in minimal supersymmetry'', 
\PRL \textbf{84} (2000) 228.

\bibitem{d0_PLB} V. Abazov {\it et al}. [D0 Collaboration], %``Search for the rare decay $B^0_s\to \mu^+\mu^-$'', 
\PLB \textbf{693} (2010) 539. 

\bibitem{cdf_PRL} T. Aaltonen {\it et al}. [CDF Collaboration], %``Search for $B^0_s\to \mu^+ \mu^-$ and $B^0_d \to \mu^+ \mu^-$ decays with 2 fb$^{-1}$ of $p\bar{p}$ Collisions'',
\PRL \textbf{100} (2008) 101802.

\bibitem{cdf_preliminary} T.~Aaltonen {\it et al}. [CDF Collaboration], %``Search for $B^0_s \to \mu^+ \mu^-$ and $B^0 \to \mu^+ \mu^-$ decays in $3.7$ fb$^{-1}$ of $p\overline{p}$ collisions with CDF II'',  
CDF Public Note 9892.

\bibitem{LHCbdetector} A.A. Alves {\it et al}. [LHCb Collaboration], %``The LHCb detector at the LHC'', 
\JINST \textbf{3} (2008) S08005.

\bibitem{paper} R. Aaij {\it et al.} [LHCb Collaboration], %``Search for the rare decays $B^0_s \to \mu^+ \mu^-$ and $B^0 \to \mu^+ \mu^-$'', 
\PLB \textbf{699} (2011), 330. 

\bibitem{Read_02} A.L. Read, %``Presentation of search results: the CL$_{\rm s}$ technique'', 
\JPG \textbf{28} (2002) 2693; \\
T. Junk, %``Confidence level computation for combining searches with small statistics'',
\NIMA \textbf{434} (1999) 435. 

\bibitem{roadmap} 
B. Adeva {\it et al}. [LHCb Collaboration], %``Roadmap for selected key measurements of LHCb'', 
arXiv:0912.4179.

\bibitem{Karlen} D. Karlen, %``Using projections and correlations to approximate probability distributions'', 
{\it Comp. Phys.} \textbf{12} (1998) 380.

\bibitem{thesis2010_068} D. Martinez Santos, %``Study of the very rare decay $B^0_s \to \mu^+ \mu^-$ in LHCb'', 
CERN-THESIS-2010-068.

\bibitem{PDG} K. Nakamura {\it et al}. [Particle Data Group],  %``Review of particle physics'', 
\JPG \textbf{37} (2010) 075021.

\bibitem{HFAG} D. Asner {\it et al}. [Heavy Flavour Averaging Group], %``Averages of $b$-hadron, $c$-hadron, and $\tau$-lepton properties'',
arXiv:1010.1589. Updated values for $f_d/f_s$ available at http://www.slac.stanford.edu/xorg/hfag/osc/end\_2009/ have been used.
%, as they include also pre-summer 2010 results which are not in \cite{PDG}.

\bibitem{BELLE_Bs} R.~Louvot, %``$\Upsilon$(5S) Results at BELLE'', 
arXiv:0905.4345.

\bibitem{TisTos} E. Lopez Asamar et al., %``Measurement of trigger efficiencies and biases'', 
LHCb-PUB-2007-073.%LHCb-2008-073.



\end{thebibliography}
\end{document}